# Can Contextualized Physics Problems Enhance Student Motivation?

Uncovering a Student-Teacher Perception Gap through a Large-Scale Survey

Yajun Wei[1,2*], Xinting Peng[2], Yi Zhong[3], Feipeng Pi[1,2], Yanfang Zhai[1,2], Lei Bao[4†]

1. Institute of Physics Education Research, Guangzhou University, Guangzhou, 510006, China
2. School of Physics and Materials Science, Guangzhou University, Guangzhou, 510006, China
3. South China Normal University, Guangzhou, 510006, China
4. The Ohio State University, Columbus, OH 43210, USA

-------------
*Contact author: Yajun Wei, Email: weiyajun@gzhu.edu.cn

†Contact author: Lei Bao, Email: bao.15@osu.edu

-------------
Contributing authors:
Xinting Peng, Email: 1582931811@qq.com
Yi Zhong, Email: 2024010163@m.scnu.edu.cn
Feipeng Pi, Email: pifeipeng@gzhu.edu.cn
Yanfang Zhai, zhaiyf@gzhu.edu.cn

**CONFLICTS OF INTEREST**：None.

## Acknowledgement

This study is supported by Guangdong Office of Philosophy and Social Sciences with grant number GD24CJY43, Guangdong Postgraduate Education Innovation Project with grant number 2024JGXM_143 and Guangdong College Students' Innovation and Entrepreneurship Training Program with grant number 202511078198.

# Can Contextualized Physics Problems Enhance Student Motivation?

## Uncovering a Student-Teacher Perception Gap through a Large-Scale Survey

**Abstract**

Embedding physics problems in real-world settings—here termed contextualized physics problems (CPP)—is widely believed to foster students' interest, motivation, and learning. However, firm evidence for this claim remains scarce. To explore this issue, we surveyed 868 secondary students and 154 teachers to examine their attitudes toward CPP and investigate whether students and teachers perceive these problems as promoting student interest and motivation in learning physics. The findings reveal a divergence between teacher and student perspectives. While most teachers view CPP as essential for enhancing interest and motivation, student responses tell a different story. Contextualized problems appear to boost interest and motivation only among 8th graders who are newly introduced to the subject. From 9th to 11th grade, students expressed a clear preference for de-contextualized physics problems (DPP) and generally disagreed that CPP increased their interest in physics. Gender differences were also observed among younger students, with boys showing a moderately stronger preference for CPP than girls. These results provide valuable insights for educators in designing course materials and creating effective test and exercise questions. The discrepancy between teacher and student perceptions, as highlighted by the surveys and interviews, underscores the need to address this gap through targeted teacher training and professional development.

# I. INTRODUCTION

Over the past twenty years, a global trend of diminishing interest, enthusiasm, motivation, and attitudes towards science among young people has been observed [1-6]. In response to this worrying trend, numerous studies have suggested instructional strategies to reignite interest in science [7-8]. Within these strategies, several researchers have highlighted the importance of contextualizing learning activities in physics [9-12]. This suggestion stems from the observation that when science is taught in a theoretical and decontextualized manner, devoid of societal relevance, it fails to engage students [13]. Most research favors the teaching of physics concepts, laws, and principles in an everyday context.

However, there is limited research on using contextualized physics problems (CPP), although they have been widely employed in tests and exercises [14]. This raises a pivotal question: Can the use of contextualized physics problems in exercises and tests, as opposed to decontextualized ones, increase students' interest and motivation in learning physics? Anecdotally, many experienced teachers and researchers believe in the effectiveness of contextualization in exercises and tests [15-16]. However, no large-scale empirical study has provided evidence to support this belief. This study seeks to explore whether student and teachers perceive contextualized physics problems as boosting students' interest and motivation.

The concept of contextualization in teaching is derived from the theory of situated cognition, which connects the acquisition of knowledge to the activities, cultures, or situations where this knowledge is utilized or learned [17]. Contextualized teaching, also known as contextualized learning or context-based learning (CBL), has emerged as a prominent pedagogical approach to addressing the disconnect between science education and students' everyday experiences. CBL emphasizes teaching concepts and skills through real-world contexts, making learning more meaningful and engaging for students [18-19].

Despite the abundant research on teaching science in everyday contexts [20], empirical studies specifically demonstrating the impact of contextualization in stimulating interest and promoting learning are very rare [21]. Furthermore, these few studies primarily focus on teaching physics concepts, and findings from concept teaching may not be directly applicable to physics problem-solving [22].

Before proceeding, it is essential to distinguish between contextualized teaching and contextualized problems. As the literature highlights, contextualized teaching represents a broad pedagogical approach that integrates real-world scenarios, student experiences, and societal issues into the entire curricular structure [23]. In contrast, contextualized physics problems (CPP), the focus of this study, are specific instructional artifacts—namely, exercises or assessment questions—that embed physics principles within a narrative or real-world scenario. Crucially, these problems can be used within various pedagogical frameworks, including traditional, decontextualized teaching practices [24]. This study does not evaluate the efficacy of holistic contextualized teaching and learning approaches but rather investigates the specific impact of using CPPs on student motivation and interest, a practice that has become widespread in testing and exercises.

### A. CBL in science education

CBL is rooted in the need-to-know principle, which highlights that students are more likely to engage with science when it connects to their lived realities and societal issues [25]. For instance, projects like Salter's Chemistry [26], Chemie im Kontext [27], and science-technology-society (STS) initiatives [28] have demonstrated that embedding scientific knowledge in authentic and relatable contexts enhances student interest, intrinsic motivation, and problem-solving skills. These benefits arise from the relevance of learning tasks and the autonomy provided to students when they interact with contextualized content. Similarly, Rayner [19] suggests that contextualization, as opposed to traditional decontextualized teaching methods, can enhance problem-solving skills and overall student performance. CBL has also gained traction in physics education, where abstract concepts often appear disconnected from the real world. Wilkinson [29] and Whitelegg and Edwards [30] argue that physics education, when devoid of real-world contexts, may alienate students, as the subject can seem overly theoretical and irrelevant to their lives. By incorporating authentic contexts—ranging from everyday phenomena to societal issues—CBL helps students see the utility of physics in solving real-world problems, fostering deeper conceptual understanding and engagement. For example, a contextualized approach to teaching Newton's laws might involve analyzing the motion of vehicles or the mechanics of sports, making physics more relatable and accessible to learners.

Empirical studies on CBL generally indicate positive effects on student interest, motivation, and cognitive engagement. Bennett et al. [31] found that context-based approaches in science improved students' attitudes and understanding, especially among learners with diverse prior knowledge levels. In a meta-analysis, Ültay and Çalık [32] highlighted CBL's success in fostering situational interest and motivation. However, the efficacy of CBL often depends on the characteristics of the context itself (e.g., its familiarity, authenticity, and complexity) and the extent to which it aligns with students' interests and prior knowledge [33-34]. Contexts that relate directly to students' daily lives tend to trigger higher situational interest compared to more abstract or laboratory-based scenarios [35]. At the same time, uncommon phenomena or surprising contexts can appeal to students with advanced prior knowledge and interest in science, suggesting a need for differentiated approaches to context selection [36]. However, as noted by Taasoobshirazi et al. [37], many studies on contextualized learning rely on anecdotal evidence or poorly designed experiments, undermining the credibility of their conclusions.

### B. Contextualized physics problems

While CBL has been extensively explored for teaching scientific concepts, there is a notable gap in research regarding its application to physics problem-solving, particularly in the form of contextualized physics problems (CPP). CPP refer to physics exercises that embed scientific principles within real-world scenarios to enhance student engagement. Contextualization can exist at different levels: (1) a minimal level, where only the object in the problem is renamed a real world one (e.g. race car) instead of an abstract model such as a point object; (2) a moderate level, where a narrative is introduced, placing the object within a real-world setting (e.g., a dragon boat race as a setting for a motion problem); and (3) a deep level, where the physics principles are fully integrated into a complex, meaningful real-world scenario, encouraging students to apply physics reasoning in authentic, problem-solving contexts. In standardized exams and textbook exercises, CPP with minimal and moderate contextualization are more widely used [38-40].

To develop a more precise understanding of CPP, we adopt a sociocultural framework that treats context not as a mere backdrop but as an integral element of learning. Finkelstein provides a particularly useful model, identifying at least three intertwined levels of context: task, situation, and idioculture [24]. The task refers to the specific problem itself (e.g., its wording, storyline, representation). The situation describes the immediate environment where the task is performed (e.g., a collaborative classroom activity versus a high-stakes, timed exam). Finally, the idioculture refers to the shared norms and beliefs of the classroom community.

This framework helps clarify the scope of our investigation. The contextualization we study—embedding physics principles in a narrative—primarily operates at the task level. Indeed, the problems used in this study, selected from high-stakes exams and textbooks, are representative of this type. For instance, a problem that reframes a standard kinematics calculation within a narrative about a dragon boat race is a classic example of manipulating context at the task level. By focusing our analysis here, we aim to isolate the perceived effects of problem formulation as it is commonly practiced in physics assessment.

Traditionally, physics exercises and assessments frequently rely on decontextualized problems (DPP) that prioritize simplified representations and equations. While this approach simplifies problem-solving, it may fail to engage students or highlight the real-world relevance of physics principles [14]. Glynn and Koballa [41] explain that contextualization helps in applying concepts and skills in real-world situations relevant to students from various backgrounds.

Recently, CPP has been extensively used in exercises and tests, but few studies have explored its effect on student learning. Park and Lee [15] investigated student preferences for CPP and found that factors such as complexity, reading load, and time constraints hindered their engagement with contextualized problems. Similarly, Bouhdana et al. [21] examined the psychophysiological impact of CPP on university students, revealing gendered differences in responses to contextualized tasks. These findings highlight the potential challenges of implementing CPP effectively, particularly in exam-driven environments where efficiency and accuracy are paramount.

### C. Research questions

In summary, despite the abundant research on teaching science in everyday contexts [20], empirical studies specifically demonstrating the impact of contextualization in stimulating interest and promoting learning remain limited [21]. Furthermore, the focus of these studies has been primarily on teaching physics concepts, which may not necessarily translate to teaching problem-solving [22].

Despite claims in the literature that contextualized physics problems (CPP) can foster student interest and motivation, empirical evidence to substantiate this assertion remains limited. This study aims to address these gaps by systematically examining the attitudes of teachers and students toward CPP and comparing their perceived benefits to those of decontextualized physics problems (DPP). To guide this investigation, the following research questions were developed:

RQ1: How do secondary school teachers perceive the relative effectiveness of contextualized physics problems (CPP) compared to decontextualized physics problems (DPP)?

RQ2: To what extent do students believe contextualized physics problems (CPP) enhance their interest and motivation in learning physics?

RQ3: How do students' beliefs about the influence of contextualized physics problems (CPP) on their interest and motivation vary with their experience in learning physics?

## II. METHODS

### A. Research design

This study employed a mixed-method design to investigate the student perception of the impact of contextualized physics problems (CPP) on their interest and motivation, and to compare these perceptions with teacher beliefs. The research process was guided by three main steps. First, teacher perceptions regarding the superiority of contextualized over decontextualized physics problems (DPP) were examined. Second, the study investigated to what extent do students believe CPP enhance their interest and motivation in learning physics compared to DPP. Third, it explored how students' beliefs about the influence of CPP on their interest and motivation vary with their experience in learning physics. Surveys were complemented with semi-structured interviews. Before conducting the study, permissions were obtained from students and their parents, school administrators, and the institution's research committee.

For the first step, a survey comprising nine items designed to measure teacher perceptions of contextualized physics problems was administered to junior middle school and senior high school physics teachers. Participants were asked to indicate their level of agreement with each statement, choosing from one of four options: strongly disagree, disagree, agree, or strongly agree. The survey was conducted using a mobile phone-based surveying technology called *Questionnaire Star* [42].

For the second step, students randomly selected from two sampling schools were each given a sheet of paper, either Type A or Type B. Each type contained eight multiple-choice physics problems. In Type A, Problems 1–4 were presented in a decontextualized format, while Problems 5–8 were presented in a contextualized format. In Type B, Problems 1–4 were presented in a contextualized format, while Problems 5–8 were presented in a decontextualized format. The purpose of having two types of papers was to ensure that each student was exposed to both contextualized and decontextualized problems while minimizing potential order effects or biases. Student performance on these problems was not assessed; instead, the problem sheet served as a reference to remind students of the two formats of physics problems—contextualized and decontextualized—before they completed the survey. After reading the problems, students were asked to complete a survey containing nine statements related to interest, motivation, and difficulty perception. These statements, shown in Table 2, were adapted from the teacher survey to reflect the student perspective while addressing the same underlying topics.

Students indicated their level of agreement with each statement, choosing from one of four options: strongly disagree, disagree, agree, or strongly agree. The student survey was administered in paper-pencil format. One reverse-coded item was included to filter out inattentive participants. At the top of the survey, a pair of problems from the 8 problems were presented side by side again to exemplify the two types of problems. Students were informed that half of the problems they had just read were contextualized and the other half were decontextualized.

This procedure was repeated for students in grades 9, 10, and 11. The collected data were analyzed and compared to explore the evolution of student attitudes as their years of learning physics increased.

Finally, semi-structured interviews were conducted with a stratified sample of students and teachers to complement and elucidate the survey findings.

### B. Sample and materials

A total of 868 secondary school students (460 boys and 408 girls), aged 13–18, participated in the study. All participants were enrolled in standard physics curricula mandated by the authority at the national level, which specifies the precise syllabus and content for each grade level. Teachers must strictly adhere to this official curriculum and utilize the standardized textbooks provided, ensuring uniformity of content coverage across different classrooms and schools. The participants were recruited from two schools located in Guangdong Province, China. The choice of schools was made through convenience sampling. Selection of students inside the schools was random. Additionally, 154 physics teachers (101 males and 49 females) participated in the teacher survey. They are based in many different provinces across the country and the survey was delivered online. The sampling strategy ensured a broad representation of secondary-level physics learners and educators.

Eight contextualized physics problems were selected from high-stakes examinations (e.g., the national college entrance exam, senior high school entrance exam, and large-scale regional mock exams). These problems had been validated by expert committees to ensure quality and curricular alignment. Students were administrated to read these questions immediately preceding the survey, intended to remind students of the difference between the two formats. Contextualization typically involved embedding the physics principle within a real-world scenario, whereas the decontextualized version removed the scenario while retaining the same core physics concept. A panel comprising four expert-level school physics teachers and two university physics education researchers collaborated to create the decontextualized forms of the selected problems. The four expert-level school physics teachers involved each has over 10 years of teaching experience, hold Master degrees and are municipally recognized senior teachers, and have more than five years of experience developing large-scale test problems. The two university physics education researchers hold PhDs and have over 10 years of research experience with established publications in physics education research.

Two example pairs of contextualized and decontextualized problems are shown in Figure 1, one from the Chinese National College Exam paper and the other from one of the most popular physics textbooks used in the U.S. that has also been translated into many languages. For example, in the first example, in the contextualized version, a dragon boat race scenario sets the stage for a uniform acceleration and deceleration motion problem. The corresponding decontextualized version focuses solely on an object undergoing uniform acceleration and deceleration, with no narrative context.

Dragon boat racing is a traditional Chinese sport with a long history. The Dragon Boat Festival is the most celebrated day for dragon boat racing activities. Dragon boat races are usually held on a long and straight river course. A dragon boat accelerates uniformly from rest starting from t = 0 until t = t₁, and then decelerates uniformly until the speed drops to zero at t = t₂, Among the following graphs showing the relationship between the displacement x of the dragon boat and time t, which one may be correct? (　)

An object moves in a straight line. It starts from rest and accelerates uniformly from t = 0 until t = t₁, and then decelerates uniformly until the speed drops to zero at t = t₂, Among the following graphs showing the relationship between the displacement x of the object and time t, which one may be correct? (　)

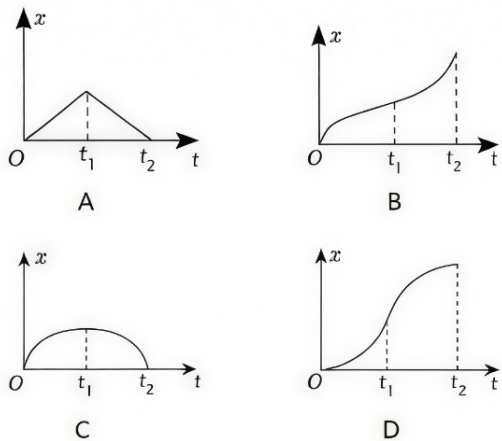

(a)

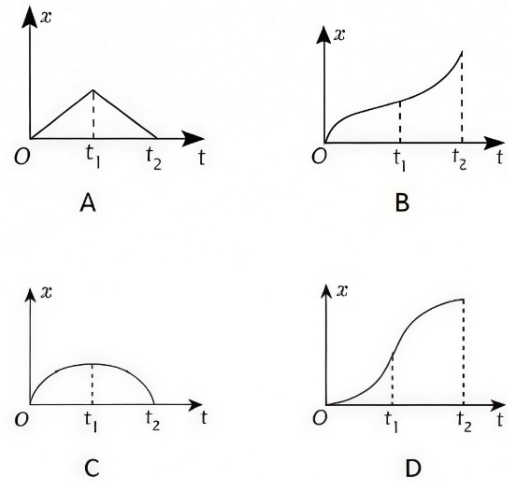

(b)

Engineers are developing new types of guns that might someday be used to launch satellites as if they were bullets. One such gun can give a small object a velocity of 3.5km/s while moving it through a distance of only 2.0cm.
(1) What acceleration does the gun give this object?
(2) Over what time interval does the acceleration take place?

An object stars from rest and reaches a velocity of 3.5km/s after traveling a distance of only 2.0cm.
(1) What is the acceleration of the object?
(2) Over what time interval does the acceleration take place?

(c)

(d)

Fig. 1. Examples of contextualized and decontextualized physics problems. A pair of contextualized (a) and decontextualized (b) problems adapted from the 2021 National College Entrance Exam in Guangdong Province, China. A pair of contextualized (c) and decontextualized (d) problems from the popular American high school physics textbook Glencoe Physics: Principles

& Problems [38].

**C. Instruments**

Two parallel surveys—one for teachers and one for students—were developed to measure perceptions of contextualized physics problems (CPP). Each survey consisted of nine items (Q1–Q9), which were adapted from prior literature (e.g., Park et al., 2004) and reviewed by experts for validity and relevance. Table 1 presents the teacher survey statements (Q1–Q9), while Table 2 presents the student survey statements (Q1–Q9), which were adapted to reflect the student perspective but retained alignment with the core topics addressed in the teacher survey.

Tab. 1. Statements to Survey Teacher Perception towards Contextualizing Physics Problems

| | |
|---|---|
| Q1 | Students are more interested in solving contextualized physics problems. |
| Q2 | Solving contextualized physics problems makes students more interested in the subject of physics. |
| Q3 | Solving contextualized physics problems makes students more motivated to learn physics. |
| Q4 | Compared to non-contextualized physics problems, students prefer solving contextualized ones. |
| Q5 | Contextualized physics problems are harder than decontextualized physics problems. |
| Q6 | Contextualized physics problems increase the amount of reading, making them more difficult. |
| Q7 | Solving contextualized physics problems can help students improve their physics grades. |
| Q8 | Solving contextualized physics problems can help students better understand physics knowledge. |

| Q9 | Physics problems should always be contextualized. |

Tab. 2. Statements to Survey Student Attitude towards Contextualized Physics Problems

| Q1 | I am more interested in solving contextualized physics problems. |
| Q2 | Solving contextualized physics problems can make me more interested in the subject of physics. |
| Q3 | Solving contextualized physics problems can make me more motivated to learn physics. |
| Q4 | Compared to non-contextualized physics problems, I prefer solving contextualized ones. |
| Q5 | I find solving contextualized physics problems harder than solving non-contextualized ones. |
| Q6 | Solving contextualized physics problems increases the amount of reading, making them more difficult. |
| Q7 | I believe solving contextualized physics problems can improve my physics grades. |
| Q8 | I believe solving contextualized problems can help me better understand physics knowledge. |
| Q9 | Physics problems should always be contextualized. |

The surveys were designed to measure two main constructs. The first construct, Motivation and Interest, was assessed using Questions 1, 2, 3, 4, 7, 8, and 9. The second construct, Difficulty Perception, was captured by Questions 5 and 6. Responses were recorded on a four-point Likert scale: 1 = strongly disagree, 2 = disagree, 3 = agree, and 4 = strongly agree. For each participant, scores were averaged within each construct. For the Motivation and Interest construct, an average score between 1 and 2.5 indicated a negative attitude towards CPP, while an average score between 2.5 and 4 indicated a positive attitude. Similarly, Difficulty Perception was represented by the average of Q5 and Q6.

The development of our survey was consistent with the Expectancy-Value model of achievement motivation [43]. This framework posits that motivation is a function of two primary components: an individual's expectancy for success and the subjective value they place on a task. Our two primary constructs map directly onto this model. The "Motivation and Interest" construct assesses the subjective value of CPPs by encompassing several related aspects. This evaluation considers the intrinsic value a person finds in the work, such as their interest and enjoyment in solving the problems (Q1, Q2). It also incorporates the utility value, which is how useful someone perceives the problems to be for their personal goals, like improving grades or understanding physics (Q3, Q7, Q8), as well as the attainment value—the importance the individual places on doing well (Q4, Q9).

Our second construct, "Difficulty Perception," relates to the "Expectancy" component, as well as the "Cost" of engaging in the task. It assesses students' perception of the cognitive demands and effort required for CPPs (Q5, Q6), which influences their expectation of success and the perceived trade-offs of tackling such problems. Grounding our survey in this framework provides a clearer conceptual basis for interpreting what is being measured and enhances the study's theoretical alignment.

To further validate the survey questions, an Exploratory Factor Analysis (EFA) was conducted on the student responses. The scree plot indicated a natural cutoff after two factors, aligning with the proposed two-construct model (Motivation and Interest, Difficulty Perception). Figure 2 presents the scree plot of the eigenvalues associated with each factor. The leveling off after the second factor supported the decision to retain two factors.

The factor loadings revealed that Q1, Q2, Q3, Q4, Q7, Q8, and Q9 primarily loaded on the first factor (Motivation and Interest), while Q5 and Q6 loaded more heavily on the second factor (Difficulty Perception). Table 3 displays the factor loadings for each survey item, confirming alignment with the intended constructs. This same factor structure was corroborated for the teacher survey data.

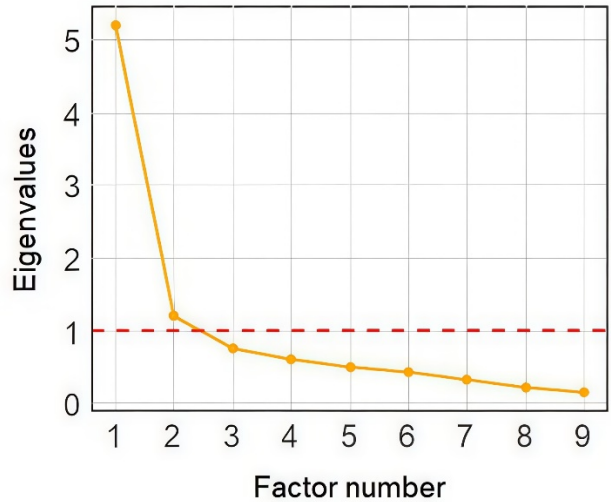

Fig. 2. Scree plot showing eigenvalues for each factor, with a cutoff at 1 (red dashed line).

Tab. 3. Factor Loadings Table

| Question | Factor1 | Factor2 |
|---|---|---|
| Q1 | 0.847 | -0.069 |
| Q2 | 0.904 | -0.148 |
| Q3 | 0.899 | -0.122 |
| Q4 | 0.825 | -0.011 |
| Q5 | -0.344 | -0.567 |
| Q6 | -0.485 | -0.755 |
| Q7 | 0.672 | -0.121 |
| Q8 | 0.741 | -0.16 |
| Q9 | 0.73 | -0.067 |

Reliability was assessed using Cronbach's alpha. For the student data, the Motivation and Interest construct (Q1, Q2, Q3, Q4, Q7, Q8, Q9) had α = 0.93, indicating strong internal consistency. Difficulty Perception (Q5, Q6) had α = 0.75, and the entire set of nine items had α = 0.79, reflecting acceptable overall reliability. For the teacher data, the Motivation and Interest construct had α = 0.89, Difficulty Perception had α = 0.76, and the entire survey had α = 0.80. These values indicate that both the student and teacher surveys are reliable and valid for measuring the intended constructs.

Following the surveys, semi-structured interviews were conducted to gather qualitative insights into the reasons behind student and teacher attitudes. A total of 12 students were interviewed, including 7 junior middle school students and 5 senior high school students, all randomly selected from the surveyed sample. Additionally, two teachers from a junior middle school and two from senior high schools were interviewed. This stratified sampling ensured a diverse range of perspectives.

Table 4 lists sample questions for student interviews, focusing on understanding the reasons behind their preferences, the perceived impact on interest and motivation, difficulty perceptions, and how their attitudes might change over time. Table 5 lists sample questions for teacher interviews, focusing on their observations and experiences with student engagement, interest, and motivation across different grade levels.

Tab. 4. Sample Student Interview Questions

| | |
|---|---|
| 1 | Do you know the difference between contextualized and non-contextualized physics problems? |
| 2 | Which kind of problem do you prefer? Why? |
| 3 | Junior middle school: Do you think this preference will change as your grade advanced in school? |

|   | Senior high school: Did this preference change as your grade advanced in school? |
|---|---|
| 4 | Do the contextualization of physics problems affect your interest and motivation in solving them? How? |
| 5 | Why do you think contextualizing physics problems increase/decrease your interest and motivation in physics problem solving? |
| 6 | Which is more difficult, contextualized/de-contextualized physics problems? Why? |
| 7 | Do you like physics ? |
| 8 | Do you think contextualized problems will motivate you to learn physics? |

Tab. 5. Sample Teacher Interview Questions

| 1 | Based on your experience, do you think contextualized physics problems engage students more effectively? Can you provide examples? |
|---|---|
| 2 | Have you received any feedback from students about the type of physics problems they prefer? |
| 3 | In what ways do you think contextualizing physics problems affects student understanding, interest and motivation? |
| 4 | Have you noticed any differences in how students of different ages or grades respond to contextualized problems? |

**D. Data analysis**

Descriptive statistics were computed to summarize survey responses. Comparisons across grade levels were conducted to assess how attitudes and preferences evolve with increased exposure to physics learning. Appropriate inferential statistical tests were employed to examine differences by grade and gender. The Exploratory Factor Analysis (EFA) and reliability analyses were performed to confirm the factor structure and consistency of the survey instruments. The confirmed factor structures guided the interpretation of student and teacher attitudes, with factor scores serving as indicators of positive or negative attitudes toward contextualized problems.

To compare teacher and student perceptions, Mann-Whitney U-tests were conducted for each survey question to determine whether significant differences exist between the two groups. The effect sizes were computed using Cohen's *d* to quantify the magnitude of the differences. Additionally, Chi-square tests were used to examine whether students and teachers exhibited a neutral or polarized stance on each survey item. For gender differences in student attitudes, U-tests were performed across grade levels to investigate whether male and female students had significantly different perceptions of CPP. To assess how student attitudes might have evolved with increased physics learning experience, trends in motivation, interest, and difficulty perception were analyzed across different grade levels using cross-sectional data.

Semi-structured interviews were conducted by a team of four graduate students in physics education research on 12 students from the two schools. The interviews were audio-recorded and transcribed verbatim. The primary purpose of the interviews was to verify and provide context for the survey findings. Interview data was also collected from four teachers from three different schools. Student and teacher responses were reviewed for alignment with the survey data, focusing on key areas such as interest, motivation, difficulty perception, and changes in attitudes over time. Particular attention was given to identifying whether interview responses confirmed the survey trends or highlighted notable exceptions. Representative quotes were selected to illustrate the perspectives of students and teachers, helping to explain observed discrepancies in the survey results. These qualitative findings served to strengthen the interpretation of the quantitative data.

## III RESULTS

### A. Comparing attitudes of teachers and students.

Figure 3 illustrates a large perceptual gap between teachers and students regarding the value of contextualized physics problems (CPP). While teachers overwhelmingly agree that contextualization enhances student interest, motivation, and learning outcomes, students exhibit a notably more negative or hesitant stance. Across several items (e.g., Q1–Q4), the majority of teachers strongly agree that contextualized problems increase engagement and interest, whereas students are far less enthusiastic and, in some cases, openly disagree. Teachers generally view it as a constructive challenge, while students perceive it as a deterrent. This divergence underscores a meaningful disconnect: the teaching community embraces contextualization as a beneficial strategy, yet the intended beneficiaries, the students, remain unconvinced of its value.

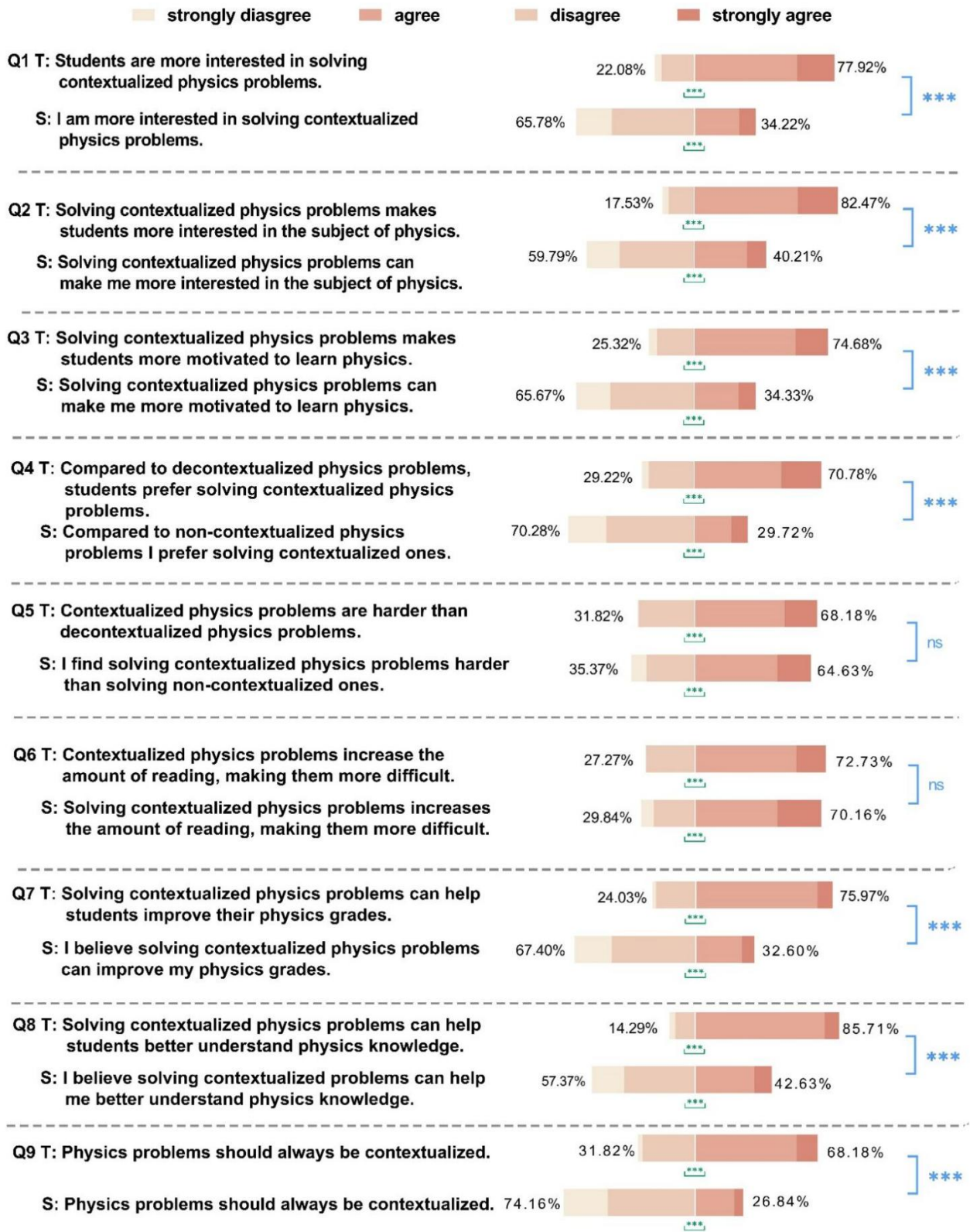

Fig. 3. Teachers (T)' and students (S)' responses to the survey questions. *** = significant

difference with $p < 0.001$, ns = no significant difference for Chi-square tests. The percentage on the left side of the bar shows the percent of 'strongly disagree' and 'disagree'. The percentage on the right side of the bar shows the percent of 'strongly agree' and 'agree.'

For students' responses to each question, a Chi-square test was conducted to check if there are statistically significant differences between those who hold positive attitude (agree and strongly agree) and those who hold negative attitudes (disagree and strongly disagree) (green horizontal comparison symbols '⌣' in Figure 3). The same analysis was performed for teachers' responses. All comparisons showed significant differences at the level of $p < 0.001$, suggesting that neither the student nor teacher population holds neutral attitudes toward any of the nine questions.

For each question, the percentage of teachers who agree (agree and strongly agree) was compared to the percentage of students who agree (vertical blue comparison symbols ']' in Figure 3). The results reveal significant differences for all the questions measuring the subconstruct of motivation and interest, indicating that teachers and students have markedly divergent perceptions of CPP's motivational effects. In contrast, no significant differences were identified between students' and teachers' perceptions regarding the subconstruct of difficulty perception. This finding suggests that while both groups acknowledge the additional cognitive demands imposed by CPP, their views diverge on its broader educational benefits. These results highlight a critical need for educators to better understand students' perspectives on CPP to bridge this perception gap effectively.

To quantify the practical significance of the differences observed between student and teacher responses, we calculated Cohen's $d$ effect sizes for each question, treating responses as a continuous 1–4 Likert scale. The results of Cohen's $d$, which measures how many standard deviations the means differ, are shown in Table 6. Significant differences with moderate-to-strong practical significance were found for questions Q1–Q4 and Q7–Q9 ($p < 0.001$, Cohen's $d$ ranging from 0.72 to 0.89). Questions Q5 and Q6 showed no significant differences ($p > 0.05$) and negligible effect sizes (Cohen's $d = 0.03\sim0.12$). These statistical findings underscore a clear perceptual gap between teachers and students regarding the educational value of contextualized physics problems.

Tab. 6. Mann-Whitney U test and Cohen's *d* effect sizes for differences between teacher and student responses

| Question | Student Mean (SD) | Teacher Mean (SD) | U statistic | *p*-value | Cohen's *d* | Effect Size Interpretation |
|---|---|---|---|---|---|---|
| Q1 | 2.24(0.88) | 2.95 (0.74) | 35353.5 | <0.001 | 0.82 | Large |
| Q2 | 2.32 (0.89) | 3.03 (0.72) | 35832.5 | <0.001 | 0.81 | Large |
| Q3 | 2.25 (0.87) | 2.88 (0.75) | 37745.0 | <0.001 | 0.74 | Medium-Large |
| Q4 | 2.19 (0.87) | 2.89 (0.79) | 36323.0 | <0.001 | 0.81 | Large |
| Q5 | 2.75 (0.86) | 2.85 (0.71) | 61476.5 | 0.375 | 0.12 | Negligible |
| Q6 | 2.86 (0.86) | 2.89 (0.66) | 64670.0 | 0.878 | 0.03 | Negligible |
| Q7 | 2.19 (0.84) | 2.83 (0.60) | 35148.5 | <0.001 | 0.80 | Large |
| Q8 | 2.34(0.89) | 2.95 (0.62) | 37716.0 | <0.001 | 0.72 | Medium-Large |
| Q9 | 2.07 (0.81) | 2.78(0.68) | 33727.5 | <0.001 | 0.89 | Large |

The semi-structured interviews with junior middle school and senior high school students, as well as their teachers, generally corroborated the patterns observed in the survey data. Several junior middle school students who preferred CPP noted that real-life contexts helped them understand the problem's meaning more easily. For instance, one 8th-grade student remarked that contextualization made the problem more interesting and understandable, although time pressure sometimes diminished this benefit. Another student commented that decontextualized physics problems seemed meaningless, suggesting that the narrative context could foster initial engagement.

However, not all younger students favored CPP. Some found that contextualization created longer texts and added complexity, making them feel overwhelmed or less willing to engage deeply. This observation aligns with the survey finding that, while 8th graders held a slightly positive or neutral attitude towards CPP, this enthusiasm was not uniform.

In contrast, interviews with senior high school students reinforced the survey's indication that older students often found contextualization less appealing. Several senior high school students explained that, while they might appreciate context outside an exam setting (finding it "interesting" or "more meaningful"), in high-pressure testing environments, they preferred straightforward, decontextualized problems that reduced reading load and complexity. This tension—context as a learning aid versus context as a cognitive burden under time constraints—mirrors the survey results showing a decline in positive attitudes toward CPP after the 9th grade.

Teacher interviews also supported the quantitative results. Junior middle school teachers noted that younger students were often drawn to contextualized problems, as the familiar settings sparked curiosity and interest. However, teachers observed that as students advanced in grade level and content difficulty increased, the same contextual details could become distracting or even intimidating. A senior high school teacher admitted that, while context-based tasks might be beneficial during concept learning stages, older students under exam pressure frequently found them cumbersome, consistent with the survey's overall pattern of declining enthusiasm for CPP over time.

### B. School type and gender influence on teacher perception

Table 6 summarizes teacher perceptions of CPP in relation to two sub-measures: "Motivation and Interest" and "Difficulty Perception". Although both junior middle school teachers and senior high school teachers favor CPP, their degrees of preference appear to differ. To determine if such a difference is statistically significant, two U-tests were conducted for the two sub-measures, with results presented in Table 7. The results indicate that junior middle school teachers are significantly more positive toward CPP than their senior high school counterparts.

Tab. 7. Comparison between junior middle school and senior high school teachers belief about contextualized physics problems

| Sub construct | Subject Group | N | Mean | Median | SD | Comparison |
|---|---|---|---|---|---|---|
| Motivation and Interest | Junior middle school | 68 | 3.07 | 3.00 | 0.463 | U-test，$p = 0.0012$, *** significant difference |
| | Senior high school | 86 | 2.77 | 2.86 | 0.563 | |
| Difficulty perception | Junior middle school | 68 | 2.75 | 3.00 | 0.614 | U-test，$p = 0.037$, * significant difference |
| | Senior high school | 86 | 2.97 | 3.00 | 0.586 | |

Table 8 summarizes teachers' perceptions of CPP by gender for the two sub-measures: Motivation and Interest, and Difficulty Perception. Two U-tests were conducted for these sub-measures, with both *p*-values larger than 0.05 (Table 8). The findings indicate no significant difference between male and female teachers for either Motivation and Interest or Difficulty Perception.

Tab. 8. Comparison between male teacher and female teacher belief about contextualized physics problems

| Sub construct | Gender | N | Mean | Median | SD | Comparison |
|---|---|---|---|---|---|---|
| Motivation and Interest | Male | 105 | 2.89 | 3.00 | 0.531 | U-test，$p = 0.565$ insignificant difference |
| | Female | 49 | 2.94 | 3.00 | 0.565 | |
| Difficulty perception | Male | 105 | 2.84 | 3.00 | 0.610 | U-test, $p = 0.443$ insignificant difference |
| | Female | 49 | 2.94 | 3.00 | 0.601 | |

### C. Evolution of student attitude

Based on the author's interaction with students of different age groups, it appears that students' attitudes towards contextualized problems may evolve with their years of physics learning experience. However, no previous research has specifically explored this topic.

To examine how the impact of contextualized physics problems (CPP) on interest evolves over time, Figure 4 plots the percentage of students who agree with the positive effect of CPP against their years of physics learning experience (represented by black dots for the mean values of the responses from male and female students).

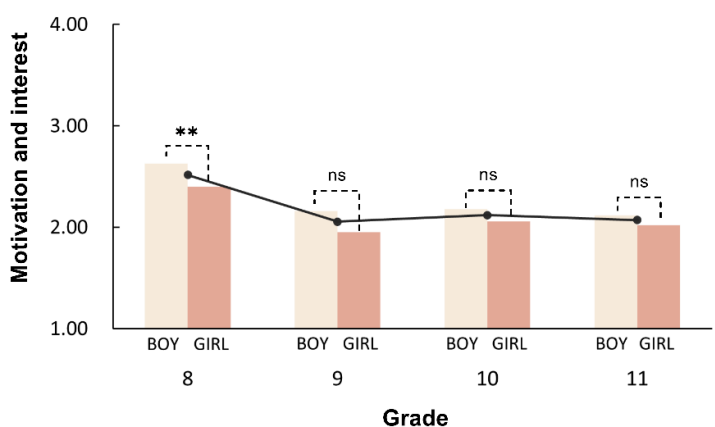

Fig. 4. Changes in students' beliefs about CPP's effect on motivation and interest across grades were analyzed. Four U-tests comparing gender differences in attitudes yielded $p$-values of 0.005, 0.131, 0.099 and 0.119 for 8th, 9th, 10th, and 11th graders, respectively. (** indicates significant difference with $p < 0.01$; ns indicates no significant difference.)

As explained in the Methods section, a higher score represents a more positive attitude towards

CPP, while a lower score indicates a more negative attitude. A score of 2.5 represents a neutral attitude. As shown in Figure 4, students who have studied physics for only 0.3 years (8th grade) tend to have a slightly positive attitude toward CPP. However, for 9th graders with over a year of physics learning experience, CPP is perceived as having no effect on promoting motivation or interest in either solving physics problems or learning the subject. The motivation and interest score plateaus after 9th grade and remains steady through 11th grade.

The lighter bars in Figure 4 represent the perception score of boys, while the darker bars represent that of girls. U-tests revealed a statistically significant difference in attitudes between boys and girls regarding the contextualization of physics problems among 8th graders. The U-test yields a *p*-value of 0.005, indicating that younger boys hold a significantly more positive attitude towards CPP than girls. However, this gender difference disappears among older students.

To better understand why younger students exhibit a slightly positive (close to neutral) attitude toward CPP in terms of its effect on motivation and interest, Figure 5 presents the responses of 8th graders to the relevant survey questions. Seven Chi-square tests were conducted to examine whether there is a statistically significant difference between the percentages of agreement and disagreement for each question. These comparisons yielded *p*-values of 0.544, 0.002, 0.182, 0.808, 0.002, 0.332, 0.002 for Q1, Q2, Q3, Q4, Q7, Q8 and Q9, respectively. The results indicate that 8th graders believe attempting CPP enhances their interest in physics, with a significant difference observed between the percentages of agreement and disagreement. Additionally, more students believe CPP may help with their understanding of physics concepts, although the difference is only slightly larger and not statistically significant. However, they also perceive that CPP negatively affects their grades. These mixed perceptions likely contribute to their overall nearly neutral attitude toward CPP.

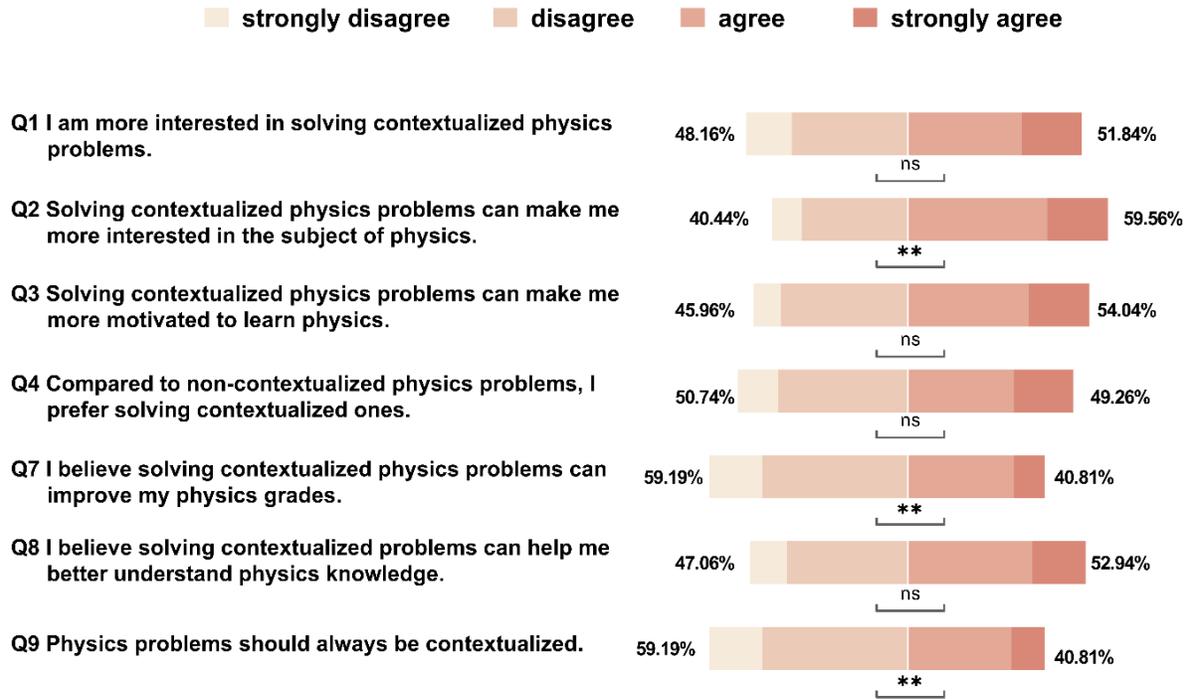

Fig. 5. Eighth-grade student responses to questions 1, 2, 3, 4, 7, 8 and 9. (Chi-square tests were conducted between positive attitudes (agree and strongly agree) and negative attitudes (disagree and strongly disagree) with ** for significant difference with $p < 0.01$ and ns for no significant difference.)

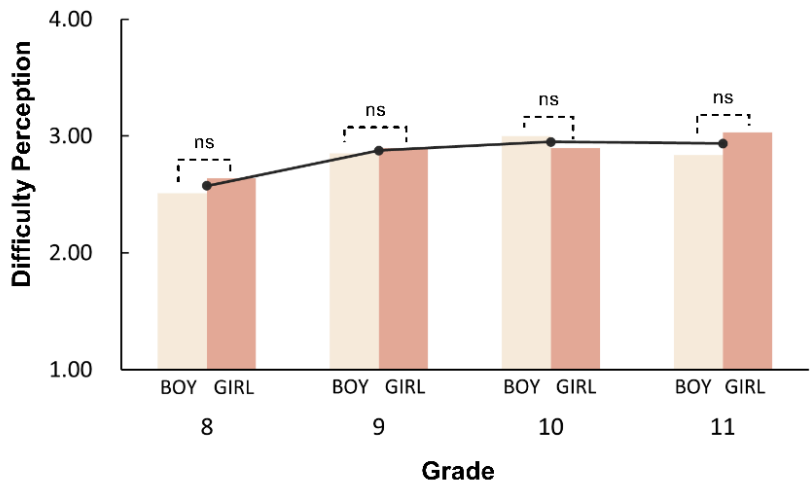

Fig. 6. Difficulty perception versus years of physics learning experience. "ns" indicates no significant difference between positive attitudes (agree and strongly agree) and negative attitudes (disagree and strongly disagree) for Chi-square tests.

Figure 6 illustrates how students' perceptions of the difficulty of contextualized physics problems (CPP) vary by grade and gender. Eighth graders do not perceive CPP as significantly more difficult than decontextualized physics problems (DPP). However, in all other grades, students strongly perceive CPP as more challenging. The perceived difficulty increases with grade level, likely because the complexity of the problems also increases, requiring more effort to extract relevant information from the context. For 8th graders, the physics problems they encounter are relatively basic, so adding context does not exceed most students' cognitive load, making the difficulty less noticeable. Four U-tests were conducted to determine whether there were statistically significant differences between boys' and girls' difficulty perceptions of CPP across all four grades. The tests yielded $p$-values of 0.211, 0.963, 0.119, 0.052 for the 8th, 9th, 10th, and 11th graders, respectively. Since all $p$-values are greater than 0.05, no significant gender differences in difficulty perception were observed.

Semi-structured interviews offer qualitative to the quantitative observations. All the interviewees confirmed that they understand the difference between CPP and DPP. Eighth-grade students who favored contextualized problems frequently mentioned that the added context made the problems more relatable and easier to understand, aligning with the survey's finding that younger learners are somewhat positively inclined toward CPP. One 8th-grade student stated, *"It helps me better understand what the problem is trying to express."* Another student noted, *"Sometimes, these problems make me more interested in physics, but if the description is too long, I feel less willing to engage with them."*

In contrast, older students consistently emphasized that as their physics curriculum became more complex, they preferred problems that presented information more directly and concisely. One high school student explained, *"When I practice physics problem solving, I like problems with real-world contexts, but sometimes a simpler problem is easier to solve efficiently."*. Another added, *"I prefer when key information is given more directly so I can focus on applying physics concepts."*.

Additionally, several senior high school students expressed an explicit preference for decontextualized problems, regardless of the situation. As one student put it, *"I prefer straightforward problems because they save time and let me focus on solving the physics part without extra distractions"*.

The teacher interviews reinforce the overall positive perception of contextualized physics problems (CPP) among educators. Junior middle school teachers were particularly enthusiastic about CPP, emphasizing its potential to engage students and make physics more relatable. One teacher stated, *"Students in junior middle school often find physics abstract, but when we connect it to real-life situations, they become much more interested and curious."*. The other teacher noted, *"Even if they struggle with calculations, the story in the problem helps them understand what's happening physically, which is important for learning physics."*. These responses align with the survey data showing that junior middle school teachers were more likely to view CPP as beneficial for student engagement.

Senior high school teachers, while still generally supportive of CPP, expressed more nuanced views. Some highlighted its role in making physics more meaningful, with one teacher commenting, *"Real-world contexts help students see why physics matters beyond exams. It's not just formulas—it's about understanding the world around them."*. One teacher noted that CPP can introduce additional cognitive load or distraction. *"Students preparing for exams tend to prefer straightforward problems. They want to get to the equations quickly without extra reading,"* he explained, *"Some students get distracted by the context rather than focusing on solving the problem, especially if the scenario is unfamiliar or overly detailed."*.

These findings reinforce the trend observed in the survey data, confirming that while younger students might initially appreciate contextualization, older students tend to favor more concise and direct problem formats. The interviews thus corroborate the survey's indication that the initial novelty and helpfulness of context diminish as students progress academically. The majority of interviewed teachers agreed that CPP has important educational value and should be incorporated into instruction and assessment despite some concerns. Their belief that CPP enhances student motivation and engagement contrasts sharply with student responses.

## IV. DISCUSSION

The core findings of this study contribute significantly to the current body of literature on the use of contextualized physics problems (CPP) in physics education. First, there is a notable mismatch between secondary school teachers' perceptions and students' experiences with CPP. While teachers believe CPP enhances motivation and interest in learning physics, students perceived a potential decrease in both motivation and interest due to the contextualization of physics problems. Second, student attitudes toward CPP become increasingly negative as their years of physics learning experience grow, shifting from a relatively positive view in the 8th grade to a strongly negative stance by the 11th grade. Third, a gender bias has been identified, with girls displaying significantly less favorability toward CPP compared to boys.

### A. Teacher perception vs. student attitude

The disparity between teacher beliefs and student attitudes towards CPP is one of the most striking findings of this study. Our results from over 150 school teachers indicate that educators overwhelmingly believe that CPP enhances student motivation and interest in physics learning. This aligns with much of the literature, which argues that embedding physics problems in real-world contexts can make learning more relevant and engaging, as highlighted in studies by Park et al. [15] and Clarke et al. [44]. Teachers, with their broader understanding of the subject, often assume that relating physics to everyday scenarios helps students grasp abstract concepts more effectively, potentially increasing both interest and performance.

However, this perception starkly contrasts with student feedback, where most students, particularly in the higher grades, report perceiving CPP as reducing their interest and motivation. This difference in perspective could be attributed to a gap between what teachers believe is pedagogically effective and how students experience these methods. For teachers, the contextualization of physics problems likely aligns with instructional goals of fostering deeper understanding and critical thinking. On the other hand, students may view these added real-world details as extraneous, distracting them from the core task of solving the physics problems. This cognitive overload, especially in an exam or time-sensitive setting, may diminish students' focus on the physics principles being tested, leading to a drop in their perceived interest and motivation. The implications of this mismatch are critical: while teachers may continue to use CPP with good intentions, they must also remain cognizant of students' responses, adjusting the balance of contextualization to ensure that it aids rather than hinders learning.

Although every interviewee could distinguish CPP from DPP when prompted, the survey itself did not supply a formal definition of "contextualization". The difference in perspective may reflect a fundamental semantic misalignment in what teachers and students understand "contextualization" to mean. Teachers, operating from a pedagogical design stance, likely interpret contextualization through a lens of pedagogical intent. For them, a CPP is a tool designed to foster authenticity, disciplinary relevance, and meaningful connections to the real world. Their positive view is rooted in this top-down understanding of the problem's purpose.

Students, in contrast, approach CPPs from a bottom-up, experiential standpoint. For them, the defining features of a CPP may not be its pedagogical intent but its surface characteristics: longer text, more complex narratives, and unfamiliar scenarios. From this perspective, "context" is not necessarily a bridge to understanding but an additional layer of information to be processed, increasing cognitive load and taking up valuable time, especially in high-stakes assessment situations. This semantic gap—where teachers see a meaningful connection and students see a complex narrative—provides a more profound explanation for our findings. It suggests the perceptual gap is not just about whether CPPs are helpful, but about what they fundamentally are to each group. This underscores the critical need for teachers to understand not only their own instructional goals but also how students perceive and define the very tools they use for learning.

This semantic gap may be understood using Finkelstein's theoretical framework of task, situation, and idioculture too [24]. While our study manipulated context at the task level (the problem's wording and story), students experience this task within the situation of a high-stakes assessment. Their negative perception is therefore a direct consequence of the interaction between the task's features (longer reading, extraneous information) and the situational demands (pressure for speed and accuracy). Teachers, in contrast, are more likely to evaluate the task in isolation, focusing on its pedagogical intent and its potential to connect with the real world, thus overlooking the powerful, and often negative, influence of the situation on the student experience. This likely explains why both groups can look at the same problem and come to entirely different conclusions about its value.

## B. Change of student attitude over time

Our finding that 8th graders generally view CPP positively, yet develop a strongly negative attitude toward it by the 11th grade, adds new insights to the existing research. Eighth graders in this study believe that CPP enhanced their interest and motivation in learning physics, a result that is consistent with prior findings by Kurbanoğlu [45], both of whom suggest that context-based questions can improve attitudes toward science and reduce test anxiety in younger students. The fact that younger students are enthusiastic about seeing real-life applications of physics is understandable—newly introduced to the subject, they find the contextualization exciting and motivating, likely because it helps make abstract concepts more concrete and relatable.

However, by the 9th grade and beyond, the excitement generated by real-world connections fades, and students begin to express a preference for decontextualized problems. This shift could be explained by the increased cognitive demands placed on students as they progress through their studies. As physics problems grow in complexity, the added context in CPP might be perceived as a burden rather than a benefit. Students may feel that contextualization unnecessarily complicates the problem-solving process, demanding extra time and effort to extract the essential information from a lengthy, context-rich scenario. Furthermore, the pressure to perform well on assessments, where speed and accuracy are critical, might lead students to prefer straightforward, decontextualized problems that allow them to focus solely on applying physics principles without distraction. The growing perception that CPP negatively affects their performance, as expressed by 8th graders who simultaneously believe CPP enhances their interest yet hampers their grades, may contribute to the overall decline in enthusiasm for CPP in higher grades.

It is important to note that our data is cross-sectional rather than longitudinal, meaning that while we observe a clear trend across different grade levels, we do not track individual students' changing attitudes over time. However, given that all students in our sample followed the same standardized physics curriculum, using identical textbooks and syllabi across grades, it is reasonable to infer that the observed trend reflects a broader shift in perception as students gain more experience in physics. Nonetheless, future research employing a longitudinal design could provide stronger evidence to determine whether individual students' attitudes genuinely change over time or if cohort-specific factors influence the results.

### C. Gender differences in CPP perception

Our study also reveals a notable gender difference in students' attitudes towards CPP. Girls are significantly more likely than boys to express a dislike for contextualized physics problems, especially in lower grades. This gender bias aligns with previous research that has explored differences in how male and female students approach problem-solving tasks, particularly in STEM subjects. Studies such as those by Murphy and Whitelegg [46] suggest that girls tend to have lower confidence in physics and are more likely to perceive science-related tasks as difficult. This could explain why junior middle school girls in our study show a stronger disfavor towards CPP, which they may perceive as adding an unnecessary layer of difficulty to an already challenging subject.

Additionally, boys' generally more positive attitude towards CPP might be attributed to their problem-solving style, which, according to Stoet and Geary [47], tends to be more analytical and tolerant of complex, multi-step tasks. However, our findings also indicate that boys' positive perception of CPP diminishes over time, though less dramatically than for girls. This could suggest that while both genders experience the added complexity of CPP as they advance in physics education, the impact is more pronounced for girls, possibly exacerbated by pre-existing confidence gaps or stereotypes about gender and physics performance.

These gender differences emphasize the importance of tailoring instructional strategies to meet the needs of both male and female students. While CPP may be an effective tool for promoting engagement and interest in younger students, educators should remain mindful of the potential gender biases that arise with increased complexity in physics problems. Addressing these biases might involve designing CPP that is more accessible or providing additional support to ensure that both boys and girls feel confident in their ability to solve contextualized physics problems.

### D. Implications for practitioners and policymakers

The disparity between teacher beliefs and student experiences with contextualized physics problems (CPP) highlights several actionable steps for educators and policymakers.

Firstly, contextualization should be tailored to students' learning stages. While younger students may benefit from CPP, its effectiveness appears to diminish as students advance in their physics education. This suggests that instead of a blanket implementation of contextualized problems, a mixed approach incorporating both contextualized and decontextualized problems may better align with students' evolving needs.

Secondly, addressing cognitive load is crucial as students progress. Advanced physics problems naturally become more complex, and the inclusion of excessive contextual details may overwhelm students. Simplifying the wording of contextualized problems or introducing scaffolded problem-solving strategies can help students focus on the core physics concepts without being distracted by unnecessary information.

In addition, supporting gender-sensitive instruction is essential. Teachers should be made aware of the observed gender differences in how CPP is perceived. Interventions such as collaborative learning, confidence-building activities, and offering diverse problem examples could help ensure both male and female students feel equally supported and engaged.

Finally, professional development for teachers is necessary to address the evolving needs of students regarding CPP. Teachers would benefit from workshops and training sessions that focus on identifying when and how to use contextualization effectively. This would enable educators to better align their instructional strategies with student feedback and learning progress, ensuring that CPP remains a valuable tool in physics education.

## V. CONCLUSIONS AND IMPLICATIONS

Through the analysis of a survey from over one thousand secondary school students and teachers, this study reveals a significant discrepancy between the perceptions of teachers and students regarding the use of contextualized physics problems (CPP). While teachers generally believe that CPP enhances student interest and motivation in learning physics, students, especially those with more experience in the subject, report the opposite, perceiving CPP as a source of difficulty rather than a tool for motivation. This misalignment becomes more pronounced as students advance in their physics education.

Additionally, gender differences are evident for younger students, with girls showing less favorability toward CPP compared to boys. These findings contribute valuable insights into the role of contextualization in physics education. The semi-structured interviews corroborate these patterns, as younger students often acknowledged that contexts initially sparked their interest, while older students expressed frustration with the added complexity and reading load.

The findings from this study suggest several important considerations for educators and policymakers. Firstly, the use of contextualized physics problems (CPP) should be tailored to the students' level of learning experience. While CPP can effectively boost interest and motivation among younger students who are new to physics, its impact appears to wane as students advance and face more complex content. Interviewees highlighted this shift, explaining that context feels less engaging and more burdensome to older students. However, this does not necessarily imply that real-world contexts should be minimized in instruction. Instead, educators should carefully scaffold the use of CPP, providing structured strategies to help students navigate the added complexity while reinforcing their problem-solving skills. Teachers should thus adjust the balance between contextualized and decontextualized problems, ensuring that context is integrated in ways that support learning rather than becoming a distraction or an additional difficulty.

Additionally, the gender disparity observed—where girls disfavor CPP more than boys—highlights the need for more inclusive teaching practices. Educators should consider offering a variety of problem types to cater to different preferences and learning styles, ensuring that all students feel engaged and supported in their physics education. Further research is warranted to explore whether tailored approaches could mitigate the perceived challenges associated with CPP.

Finally, there is a clear disconnection between teachers' perceptions and students' attitudes toward CPP, suggesting that professional development is needed to help teachers better understand how students at different stages of learning experience contextualized problems. Rather than assuming that contextualization is always beneficial, educators should be equipped with strategies to differentiate when, where, and how to use CPP most effectively. Teachers and examiners may benefit from training on how to more effectively adapt their instructional strategies and exercise/exam material based on student feedback, learning progress, and the evolving attitudes revealed by this study.